\documentclass{ptapap}

\author{Aleksandra Solarz}[NCBJ]
\author{Maciej Bilicki}[ Leiden, NCBJ, ZG]
\author{Agnieszka Pollo}[NCBJ, OAUJ]
\affil[NCBJ]{Narodowe Centrum Bada\'n J\k{a}drowych\\
  ul. Andrzeja So\l tana 7, 05-400 Otwock, Poland}
\affil[Leiden]{Leiden Observatory, Leiden University, the Netherlands}
\affil[ZG]{Janusz Gil Institute of Astronomy, University of Zielona Góra, Poland}
\affil[OAUJ]{Obserwatorium Astronomiczne Uniwersytetu Jagiello\'nskiego}

\title{Search for unusual objects in the WISE Survey}

\begin{document}

\maketitle

\begin{abstract}
Automatic source detection and classification tools based on machine learning (ML) algorithms are growing in popularity due to their efficiency when dealing with large amounts of data simultaneously and their ability to work in multidimensional parameter spaces. 
 In this work, we present a new, automated method of outlier selection based on support vector machine (SVM) algorithm called one-class SVM (OCSVM), which uses the training data as one class to construct a model of 'normality' in order to recognize novel points.
 We test the performance of OCSVM algorithm on \textit{Wide-field Infrared Survey Explorer (WISE)} data trained on the Sloan Digital Sky Survey (SDSS) sources. Among others, we find $\sim 40,000$ sources with abnormal patterns which can be associated with obscured and unobscured active galactic nuclei (AGN) source candidates. We present the preliminary estimation of the clustering properties of these objects and find that the unobscured AGN candidates are preferentially found in less massive dark matter haloes ($M_{DMH}\sim10^{12.4}$) than the obscured candidates ($M_{DMH}\sim 10^{13.2}$). This result contradicts the unification theory of AGN sources and indicates that the obscured and unobscured phases of AGN activity take place in different evolutionary paths defined by different environments. %is consistent with the evolutionary scenario 
%We calculate the angular correlation function for a sample of ~3 000  selected from the which are present in the optical surveys with median photometric redshift of $<z>=0.53$.  

\end{abstract}
\section{Introduction}
The increasing amount of data that is collected from large digital sky surveys, now reaching several peta-bytes including hundreds of million of celestial objects and thousands of parameters measured for each of the observed sources, forces astronomy into finding new ways of efficient detection, segregation and classification of the collected information.

An additional role that they will play is allowing the astronomers to search for some rare or even new astrophysical objects which were otherwise missed within the surveys. This aspect can be studied by exploring previously uncharted parts of the parameter spaces, like the classical color-color (CC) diagrams, where the distribution of already known sources can point to rare outliers. However, more often than not, new sources can hide their existence by mimicking the appearance of the regular sources. With the new, automated methods offered by Machine Learning algorithms it is now possible to work in high-dimensional parameter spaces not only to efficiently create samples of regular sources in large amounts of data but also to search for undersampled or even new celestial objects. 
The presented work is aimed at detecting anomalies within the Wide-field Infrared Survey Explorer (WISE, \citealt{wright10}) data set based on a pure training sample of galaxies, stars and quasars selected from the cross-match between WISE and Sloan Digital Sky Survey (SDSS, \citealt{york00}) catalogs.  
To find novel sources  we use \textit{domain-based novelty detection} method, which is designed to create a boundary based on the structure of the training data set SVM. More commonly, in terms of the usage of SVM for novelty detection, it is known as \textit{one-class SVM} (OCSVM).
\section{The Data}
 The WISE telescope, launched by NASA in December 2009, scanned the whole sky  in four passbands ($W1$---$W4$) covering near- and mid-IR wavelengths centred at 3.4, 4.6, 12 and 23 $\mu$m, respectively. 
Exploration of the publicly available AllWISE catalog \citep{cutri13}, which contains over 747 million sources with photometric information, allows us to test the power of basic artificial intelligence algorithms for anomaly detection in order to obtain information about special objects contained within the dataset. 
To create the training set, which is the basis of any supervised machine learning problem, we need to manually classify a representative subset of the data. 
For that purpose we performed a 1'' radius cross-match between AllWISE dataset with the
SDSS DR13 (\citealt{sdssdr13}). %, henceforth refered to as AllWISE$\times$SDSS.
This procedure resulted in 2.6 million common sources out of which galaxies comprise 74\%, quasars 13\% and stars 13\% of the sample.
The second step of data preparation for a machine learning procedure is to create a feature vector for each training example, which contains discriminating properties for an object.
To that aim we decided to use the $W1$ magnitude measurement, $W1-W2$ color and a concentration parameter $\mathtt{w1mag13}$ defined as the difference between flux measurements in two circular apertures in the $W1$ passband in radii equal to 5.5'' and 11.0'' centered on a source (previously used by, e.g., \citealt{kurcz2016}). 
\section{Method}
One of the most popular schemes used for source classification is the Support Vector Machine  (SVM, \citealt{vapnik}). The basic idea behind SVM is that the algorithm is supposed to learn to recognize two (or more) types of objects based on the training examples provided by the supervisor. 
It uses kernel functions to map the input parameter space into a higher dimensional feature space, where it will search for the best separation hyperplane between the examples of the training points from each category with the biggest margin possible. Then the remaining sources, whose nature is unknown, will have their class assigned based on their relative position to that boundary. 
%However, it cannot deal with the data from the general set with patterns unseen during the training process, which results in contaminated output samples of celestial objects.
 It is possible to modify the SVM algorithm as a detection tool for unrecognizable patterns within the data: instead of using multiple training classes of sources the user has to specify only one class, composed of all the known sources. Then, instead of creating a separation plane, the algorithm will create an enclosed hypershape containing all the known points within the feature space. When the user will apply the remaining unknown sources, all points falling outside of that hypershape will be considered as \textit{anomalies}. This modification is referred to as One-Class SVM (OCSVM) and is perfectly suited for purposes of searching for unusual or unknown sources within large astronomical datasets. For details we refer the reader to \citet{solarz17} and references therein.
\section{Results}

\begin{figure} 
 \centering
  \begin{minipage}{0.48\textwidth}
   \includegraphics[width=\textwidth]{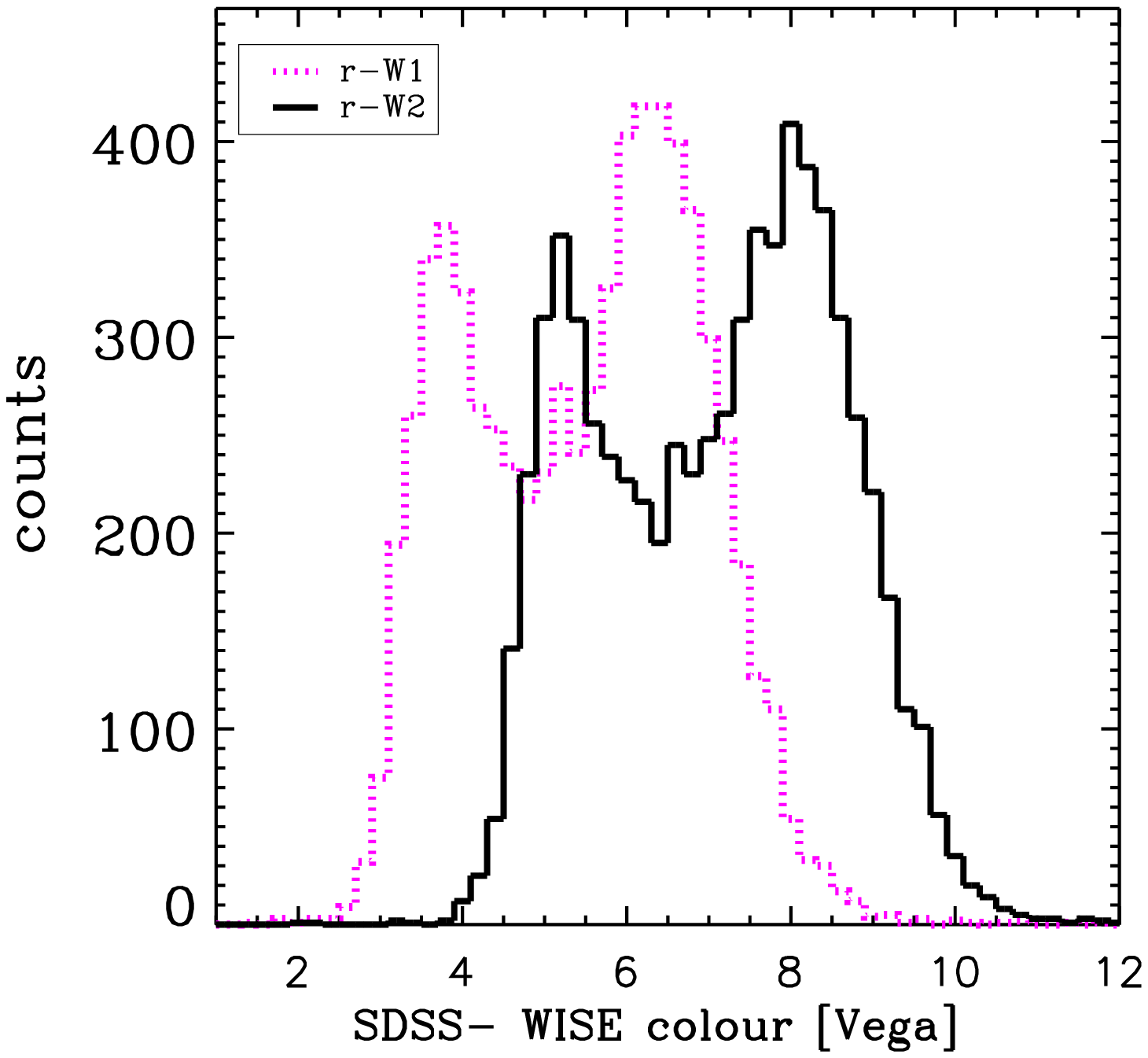}
    \caption{    \label{bimod}
Optical-infrared color distribution of OCSVM-selected anomalous WISE sources with photometric counterparts in SDSS DR14 database. Solid and dotted lines represent the $r-W2$ and $r-W1$ color distributions. Both colors show similar bimodal behavior indicating existence of at least two source populations within the OCSVM-selected data.}
  \end{minipage}
  \quad
  \begin{minipage}{0.48\textwidth}
    \includegraphics[width=\textwidth]{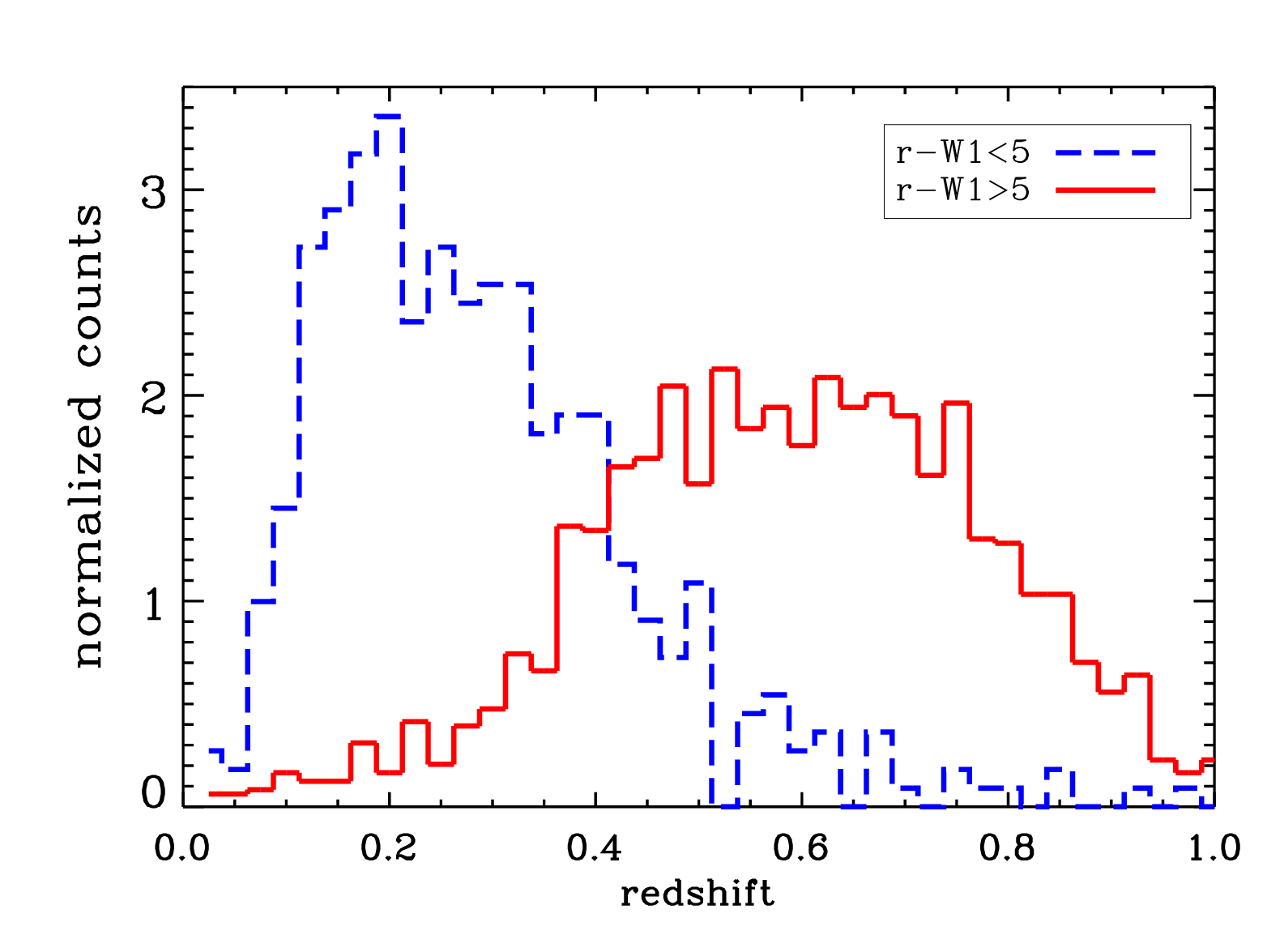}
    \caption{    \label{zdist}
Photometric redshift distribution for OCSVM-selected AGN candidates divided according to the $r-W1\sim 5$ criterion into obscured and unobscured sources.}
  \end{minipage}
\end{figure}

After training the OCSVM algorithm on the AllWISE$\times$SDSS training sample, the full AllWISE data were tested against the created normality model. As a result, we found 
 $\sim$40,000 sources showing novel properties. The distinguishable property of these sources is their extremely red $W1-W2$ color (as large as $\sim 2$ in the Vega system), which means that the sources experience a sharp increase of observed flux with the increase of the observational wavelength.
Such behavior and large mid-infrared fluxes can be associated with either warm dust emission or policyclic aromatic hydrocarbon emission lines (characteristic for star-forming galaxies). To confirm the nature of the selected anomalies we performed a positional cross-match with other publicly available data sets (irrespective of the observational wavelength). We found $\sim 7,000$ counterparts in the photometric part of the SDSS survey, meaning that these sources have optical fluxes measured through the five optical filters, but no spectroscopic redshift information is available. 
Nevertheless, about $\sim 2,700$ of these sources have their photometric redshifts estimated by \citet{beck16}. The optical-infrared color distribution shows clearly bimodal behavior indicating that at least two populations of extragalactic sources are contained within this group (see Fig.~\ref{bimod}). Similar properties were reported for obscured and unobscured active galactic nuclei (AGN) sources by \citet{donoso14}.

 Two basic types of AGNs, obscured (type-II), and unobscured (type-I), that are being widely observed, are thought to be the result of the orientation of a dust torus around the central black hole. On the other hand, the obscuration of the AGN may rise from larger dust structures like those predicted for major mergers of galaxies (e.g. \citealt{hop06}). 
Simulations by \citet{hop08} suggest that the dust obscuration could represent a phase of galaxy evolution when a central black hole cannot produce enough accretion luminosity to eject the surrounding material. 
One of the tests which can provide an answer to this problem is the measurement of the obscured and unobscured AGN clustering as it allows for measurement of the mass of the parent dark matter halo (DMH). If the unification theory is correct, then the two AGN types should appear in similar environments (i.e. similarly massive DMHs).
To test this theory we perform a clustering analysis of the two types of AGN sources found through OCSVM analysis; we divide the sample into obscured and unobscured AGNs based on $r-W1\sim 5$ criterion (cf. Table~\ref{tabelka}). To estimate the angular correlation function we used objects appearing in the northern hemisphere only, as the SDSS coverage is much larger there, and therefore the number of sources available for clustering measurements is greater.
The OCSVM-selected AGN candidate sample is not based on any spectroscopic data, only photometric information is available -- for that reason the sources used in this work have never before been used for measurements in the large scale structure context.

\begin{table}
\caption{Summary of obtained correlation function parameters.}
\label{ajaj}
\begin{center}
\begin{tabular}{|c||c|c|c|c|c|}
\hline\hline
&$N_{obj}$&$\gamma$&$r_{0}$~[Mpc $h^{-1}$]&$b$&$M_{DMH} [M_{\odot}~h^{-1}]$\\\hline
$r-W1<5$&743&$1.79\pm 0.06$&$4.57\pm 0.42$&$1.13\pm 0.10$&$10^{12.43}$\\
$r-W1>5$&1212&$1.87\pm 0.08$&$6.96\pm 0.55$&$1.98\pm 0.13$&$10^{13.20}$ \\ \hline\hline
\multicolumn{6}{c}{ }
\label{tabelka}
\end{tabular}
\end{center}
\end{table}

\begin{figure}
  \centering
  \begin{minipage}{0.48\textwidth}
    \includegraphics[width=\textwidth]{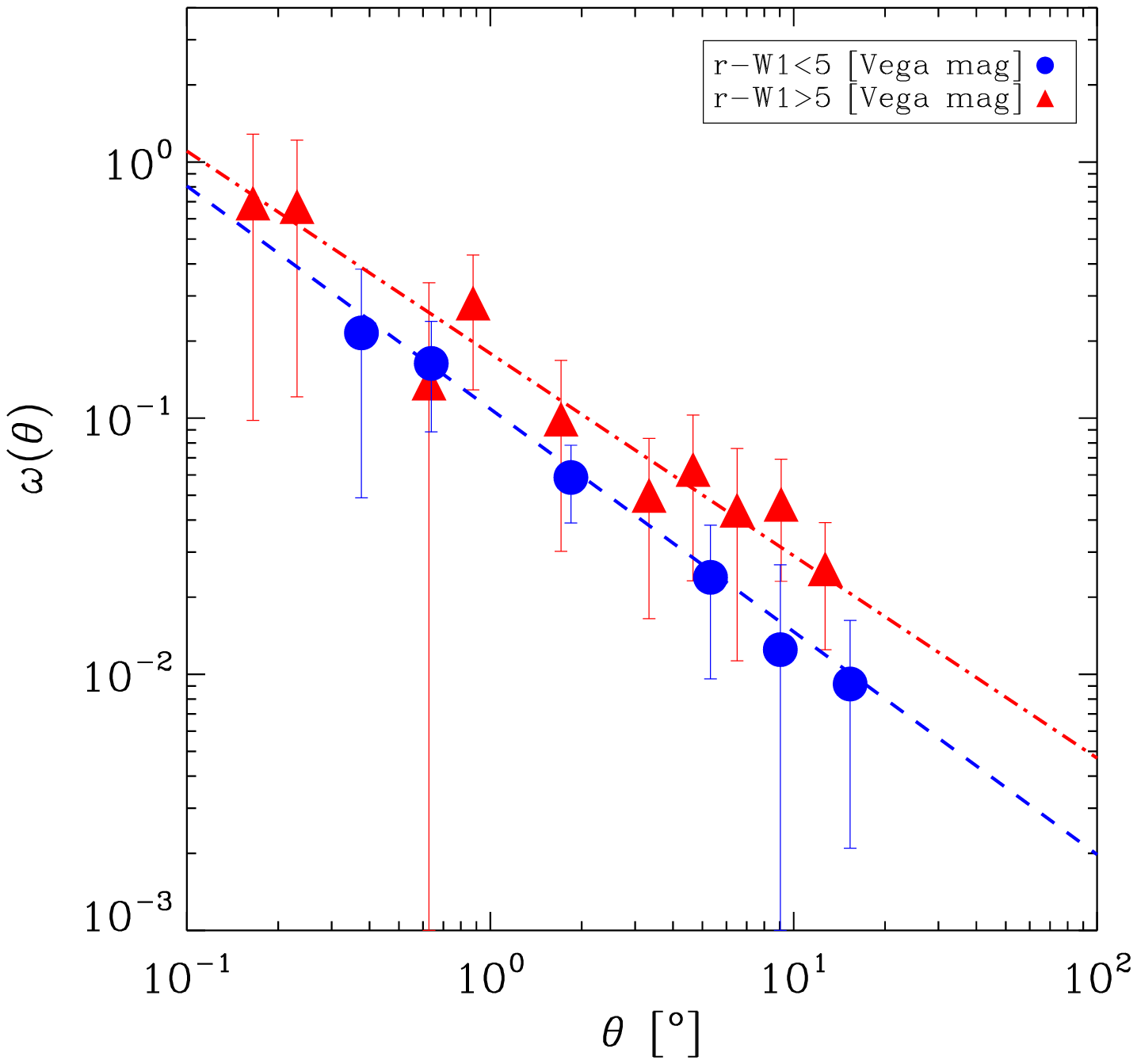}
    \caption{    \label{cf}
Resultant angular correlation function for two AGN samples: obscured (marked by red triangles) selected by $r-W1>5$ [Vega mag] and unobscured sources (marked by blue circles) selected by $r-W1<5$ [Vega mag] cuts. Dashed and dash-dotted lines represent the power-law fit to the correlation function.}
  \end{minipage}
  \quad
  \begin{minipage}{0.48\textwidth}
    \includegraphics[width=\textwidth]{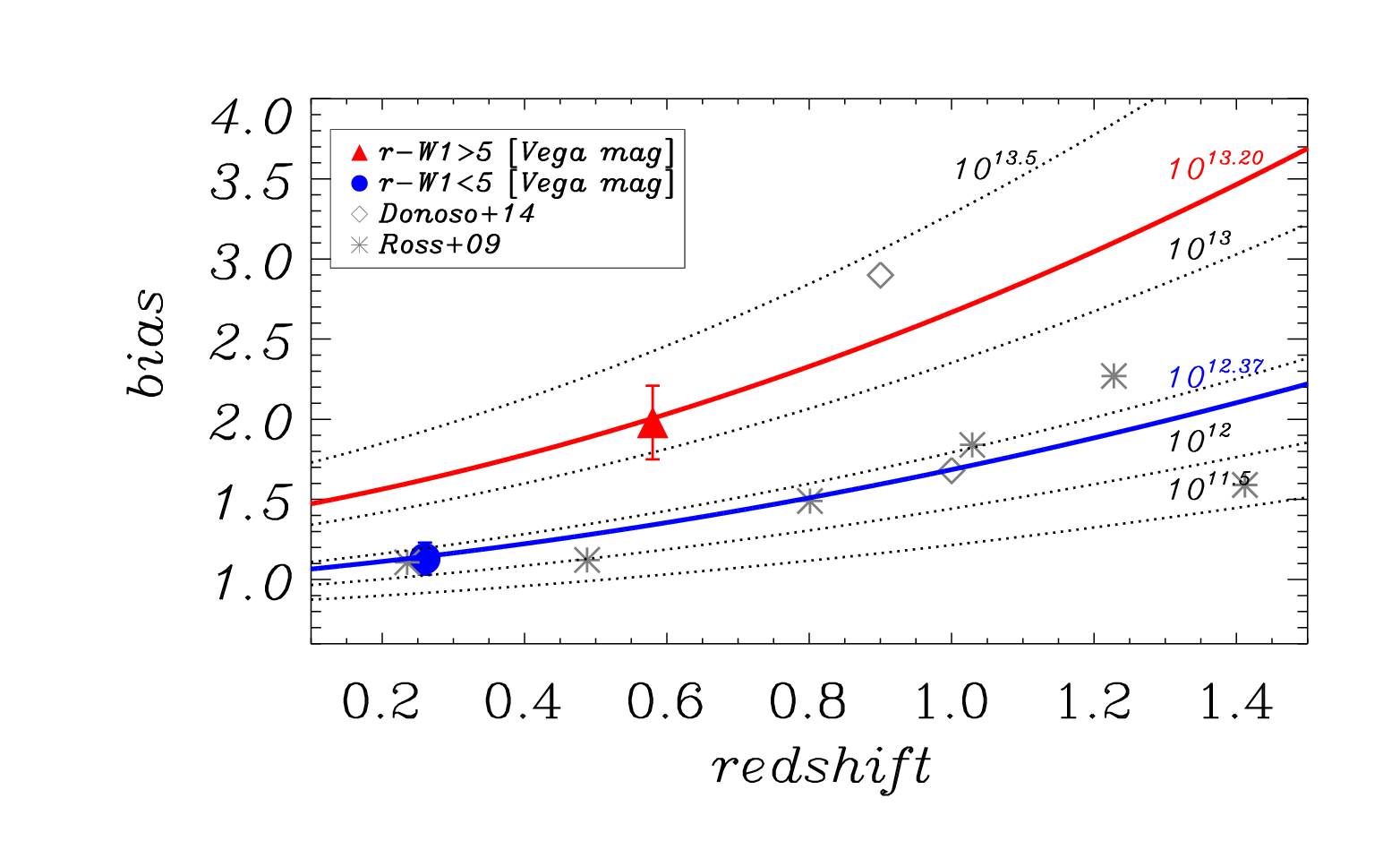}
    \caption{    \label{bias}
Linear  bias  as  a  function  of photometric redshift  for  obscured (red circle) and unobscured (blue circle) OCSVM-selected AGN candidate samples. Dashed curves represent the theoretical linear halo bias evolution of dark matter halos of minimal masses from $10^{11.5}$ to $10^{13.5}$ (bottom to top).
 As a reference we show results from the literature: diamonds from \citet{donoso14}, asterisks from \citet{ross09}.}
  \end{minipage}
\end{figure}

 We used the \citet{ls} estimator to evaluate the angular 2-point correlation function and used jack-knife resampling of 32 subsamples to evaluate the errors using full covariance matrix modeling. 
Usually a correlation function follows the power-law $\omega(\theta)=A_{\omega}\theta^{1-\gamma}$, where $A_{\omega}$ is the measurement of the correlation strength and $\gamma$ indicates its scale dependence. Using the measurements of the angular clustering, we can infer  the  3-dimensional  clustering  properties  based  on  the known  redshift distribution (shown in Fig.~\ref{zdist})  via  Limber’s  equation \citep{limber}.
The obtained results are presented in Fig.~\ref{cf} for obscured and unobscured AGN candidates.
 Then, to relate the source clustering to dark matter clustering, it is possible to use a bias parameter - a quantity which describes the differences between the clustering of baryonic field and the underlying mass distribution, i.e. $b^{2}(r,z,M)=\xi_{g}(r,z,M)/\xi_{m}(r,z)$, where $\xi_{g}(r,z,M)$ is the correlation function of the investigated source population and $\xi_{m}(r,z)$ is the dark matter correlation function. %, , or in terms of assumed mass fluctuations bias parameter should satisfy the relation $\sigma_{(g,R)}=b\sigma_{(m,R)}$. %Bias parameter depends on scale $(r)$, redshift ($z$) and mass of the objects ($M$).
For the details of the calculations we refer the reader to \citet{peebles} and referenced therein.
In Fig.~\ref{bias} we  show  the  linear  bias  evolution  derived  from \citet{sheth} formalism for varying minimum DMH mass  thresholds.
%Following Quadri07 we compute $\sigma_{R}$ for representative scale $R=8$ $h^{-1}$~Mpc. To compute the mass fluctuations we use $\sigma_{m,8}(z)=\sigma_{(m,8)}(0)D(z)$, where $0$ marks the mass fluctuation at present day...\textbf{CZY TO WSZYSTKO PISAC???}
%\begin{figure} 
% \centering
%  \begin{minipage}{0.48\textwidth}
%   \includegraphics[width=\textwidth]{w12_outin1.eps}
%    \caption{Distributions of the $W1- W2$ colour for AllWISE anomalies found in this work (solid black) compared with known sources from AllWISE$\times$SDSS used to train the OCSVM algorithm. Galaxies (1 827 241 objects) are marked by blue dashed lines; stars (298 269 objects) by orange dotted lines; quasars (141 494 objects) by magenta dot-dashed lines.}
%    \label{w12dist}
%  \end{minipage}
%\end{figure}

We find that the OCSVM selected samples of obscured and unobscured AGN candidates reside in different environments: while the unobscured AGNs at $\langle z_{phot} \rangle \sim 0.26$ are found in haloes which in the present-day Universe reach $log(M/M_{\odot}h^{-1})\sim 12.47$, the obscured sources at $\langle z_{phot} \rangle \sim 0.56$ inhabit haloes of today's
  $log(M/M_{\odot}h^{-1})$ $\sim 13.20$. The unobscured AGN halo mass is in excellent agreement with the previous works of \citet{donoso14} (for obscured and unobscured AGN found in WISE$\times$COSMOS surveys)  and \citet{ross09} (for SDSS optical quasars), who report that  $log(M/M_{\odot}h^{-1})\sim 12.3$. This difference could be a result of the flux-limited nature of the source selection: sources appearing at higher redshifts must be intrinsically brighter to appear within the detection limit of the survey. Objects with higher luminosity are found to have stronger clustering signal than the faint ones (e.g. \citealt{zehavi11}), which could explain the varying DMH masses between the obscured and unobscured AGN sources. % are found at higher redshifts than the unobscured ones, which means that the sources are intrinsically brighter, and therefore more strongly clustered (e.g. \citealt{zehavi11}).
On the other hand, the merger-driven evolutionary scenario assuming that the AGN obscuration is preceding the unobscured phase of the AGN evolution could explain the fact that the obscured AGN are preferentially found in denser environments than unobscured ones. These findings are contradictory to the AGN unification theory which assumes that the difference between the two phenomena is based solely on the orientation of the dusty torus.

\acknowledgements{This publication makes use of data products from the Wide-field Infrared Survey Explorer, which is a joint project of the University of California, Los Angeles, and the Jet Propulsion Laboratory/California Institute of Technology, funded by the National Aeronautics and Space Administration. 
AS has been supported by National  Science  Centre grant number UMO-2015/16/S/ST9/00438, MB and AP by UMO-2012/07/D/ST9/02785.
}

\bibliographystyle{ptapap}
\bibliography{solarz}

\end{document}